\documentclass[a4paper,11pt]{article}
\usepackage{jcappub}
\usepackage{amsmath}
\usepackage{amssymb}
\usepackage{graphicx}

\title{The quantum cosmological tilt and the origin of dark matter}
 
 \author[a]{C\'esar G\'omez,} 
\emailAdd{cesar.gomez@uam.es}

\author[b,c]{Raul Jimenez} 
\emailAdd{raul.jimenez@icc.ub.edu}

\affiliation[a]{Instituto de F\'{i}sica Te\'orica UAM-CSIC, Universidad Aut\'onoma de Madrid, Cantoblanco, 28049 Madrid, Spain.}
\affiliation[b]{ICC, University of Barcelona, Marti i Franques 1, 08028 Barcelona, Spain.}
\affiliation[c]{ICREA, Pg. Lluis Companys 23, Barcelona, E-08010, Spain.}

\abstract{A promising candidate for cold dark matter is primordial black holes (PBH) formed from strong primordial quantum fluctuations. A necessary condition for the formation of PBH's is a change of sign in the tilt governing the anomalous scale invariance of the power spectrum from red at large scales into blue at small scales. Non-perturbative information on the dependence of the power spectrum tilt on energy scale can be extracted from the quantum Fisher information measuring the energy dependence of the quantum phases defining the de Sitter vacua. We show that this non-perturbative quantum tilt goes from a red tilted phase, at large scales, into a blue tilted phase at small scales converging to $n_s=2$ in the UV. This allows the formation of PBH's in the range of masses $\lesssim 10^{20} gr$.}

\begin{document} 
\maketitle

\section{Introduction}

It is well established that dark matter exists. One of its most surprising properties is that it does not ``shine" but rather has negligible/none  interactions in the electroweak sector. This is an interesting aspect as, if it was luminous, it would significantly suppress the formation of large scale structures in the Universe.
(see e.g. the reviews in Ref.~\cite{Bert1,Bert2} and references therein). This matter that only manifest itself via its gravitational interaction, accounts for almost $30$\% of the total energy budget of the Universe. Unveiling the nature of this form of matter has been the subject of intense research and debate for the past 50 years. There are two basic schools of thought. Those more inspired by high energy physics that looks for a solution in extensions of the standard model and the most traditional that looks for a solution within the dynamics of gravity itself. Among this second group we can distinguish between classical and quantum approaches. The most classical approach is some form of modified classical gravity able to fit the galaxy rotation curves. The approach that we can denote as roughly quantum tries to find the origin of dark matter in the quantum gravitational properties of de Sitter like solutions to GR equations. Among this second approach to the dark matter problem, the most conservative and also the one requiring minimal input is the one originated by Zeldovich \& Novikov~\cite{ZelNov} that envisions dark matter, (more precisely cold dark matter CDM), as composed of so-called primordial black holes formed not by the standard collapse of celestial massive bodies but as a consequence of specially intense primordial quantum fluctuations, naturally appearing in an early phase dominated by early dark energy as the one needed for standard inflation. 

This line of thought that in its naive version leads to too massive black holes, useless to account for dark matter, was revitalized by Hawking \& Carr~\cite{CarrHawking,Hawking} and has suffered from the first moments several different and recently (after LIGO events, see e.g. Ref.~\cite{Kamion} as a first example, although the literature on LIGO and PBHs association is very large and by no means we can be exhaustive here) very promising avatars. It is not the goal of this brief note to go through the different inflationary models that can give rise to a reasonable description of CDM as composed of primordial black holes (see e.g. a recent review in Ref.~\cite{CarrReview}). So we focus, in a way that clearly indicates our narrow scope on the richness of models, to identify the basic, model independent needed ingredients, to develop realistic models of CDM as PBH's.

The basic ingredients are basically two:

\begin{enumerate}
\item A model of inflation with two well defined regimes for low and large comoving momentum $k$; in the low $k$ the power spectrum being consistent with cosmic microwave background (CMB) and large scale structure (LSS) experiments. For large $k$ there is enhanced power spectrum with a sort of phase transition regime between the two phases at some critical value $k_c$.
\item This change in the behavior of the power spectrum of curvature fluctuations is normally associated with a change in the nature of the tilt measuring departures from scale invariance. Namely, the phase at large $k$ where we expect the enhancement of the power spectrum needed to trigger PBH formation is characterized by a {\it blue tilt}. This blue tilt is what makes the evolution of the power spectrum after the critical value $k_c$ to increase in the form required by PBH formation.
\end{enumerate}

Within the standard inflationary approach, the two conditions above can be manufactured  in models with two inflationary fields, that generically we can denote {\it inflaton} and {\it curvaton}\footnote{Among the different inflationary models accounting for a blue tilt regime (see Ref.~\cite{Kuhnel} for a more complete list of references) we will use as guiding model the curvaton scenario~\cite{curvaton}. For an approach motivated by hybrid inflation see e.g. Ref.~\cite{bellido}}. The power spectrum at small $k$ should be dominated by the inflaton power spectrum we observe in CMB experiments with an absolute value at CMB scales of order $10^{-9}$ and a scale dependence dominated by a red tilt with spectral index $n_s\sim0.96$. For large values of $k$ the curvaton field, in order to account for PBH formation, should control the dynamics leading to a curvaton power spectrum with {\it blue tilt} that creates density perturbations at large $k$ of order (see e.g. Ref.~\cite{Kawasaki:2012wr})
\begin{equation}\label{one}
\sqrt{{\cal{P}}_{\rm curvaton}} \sim \frac{\delta \rho}{\rho} \sim 10^{-3/2}
\end {equation}
instead of the one defined at CMB scales
\begin{equation}
\sqrt{{\cal{P}}_{\rm inflaton}} \sim \frac{\delta \rho}{\rho} \sim 10^{-9/2}
\end {equation}
Assuming primordial Gaussian density perturbations, the constraint (\ref{one}) leads to a value of the $\beta$ parameter measuring the ratio of PBH density to total density
\begin{equation}
\beta \sim \sqrt{{\cal{P}}_{\rm curvaton}} e^{-\frac{1}{18{\cal{P}}_{\rm curvaton}}}
\end{equation}
where we take the variance of the density contrast to be of the same order as the curvaton power spectrum. This constraint is (as shown in Ref.~\cite{Kawasaki:2012wr})
compatible with PBH in the window of masses around $10^{20}gr$.

The crucial question is how to get naturally these two phases with the appropriate hierarchy and without artificially increasing the degrees of freedom by adding extra fields to the inflaton potential. In this note we present a tentative {\it model independent} answer to this question. 

\begin{figure}
\centering
\includegraphics[width=.88\columnwidth]{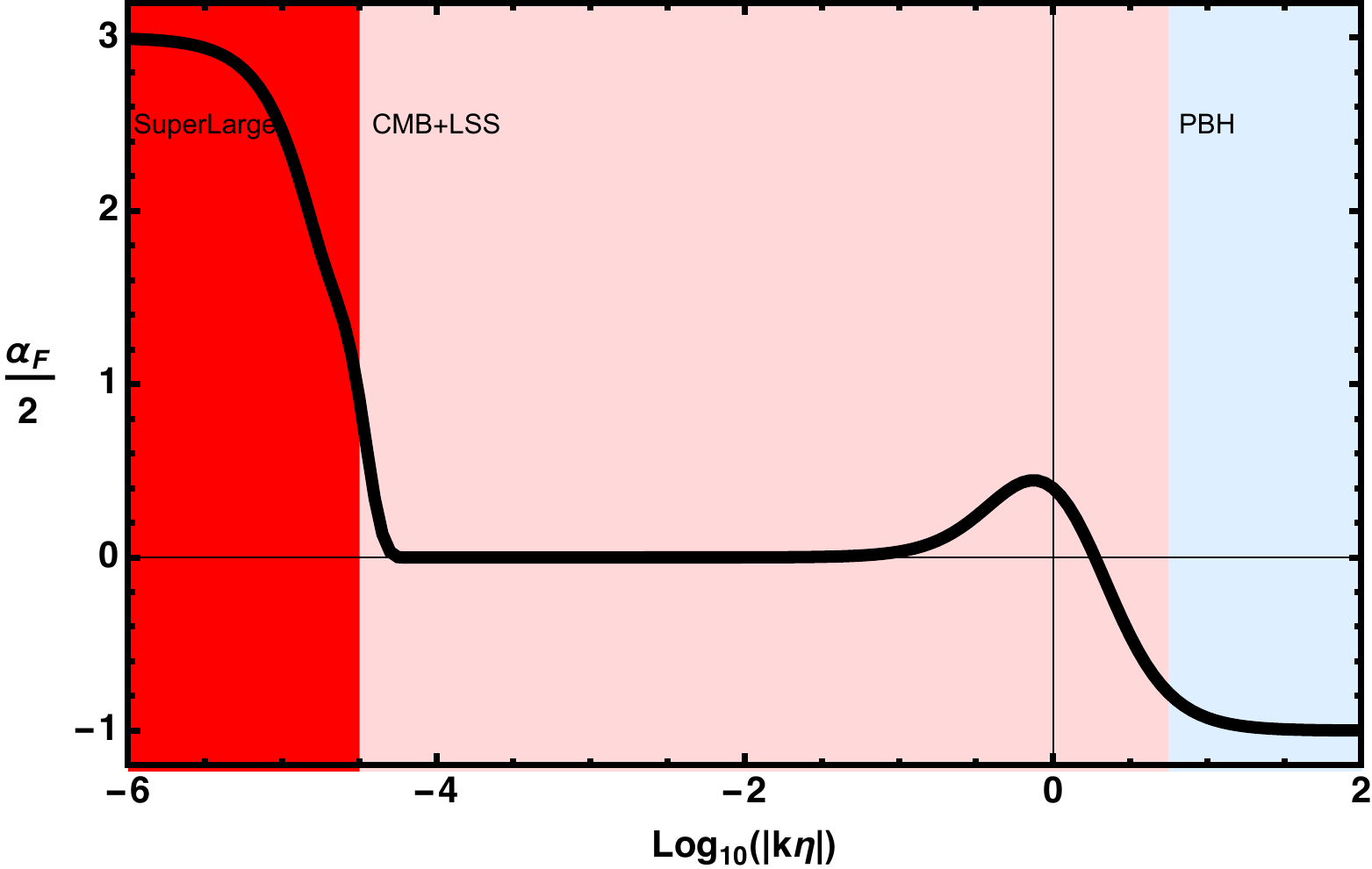} 
\includegraphics[width=.89\columnwidth]{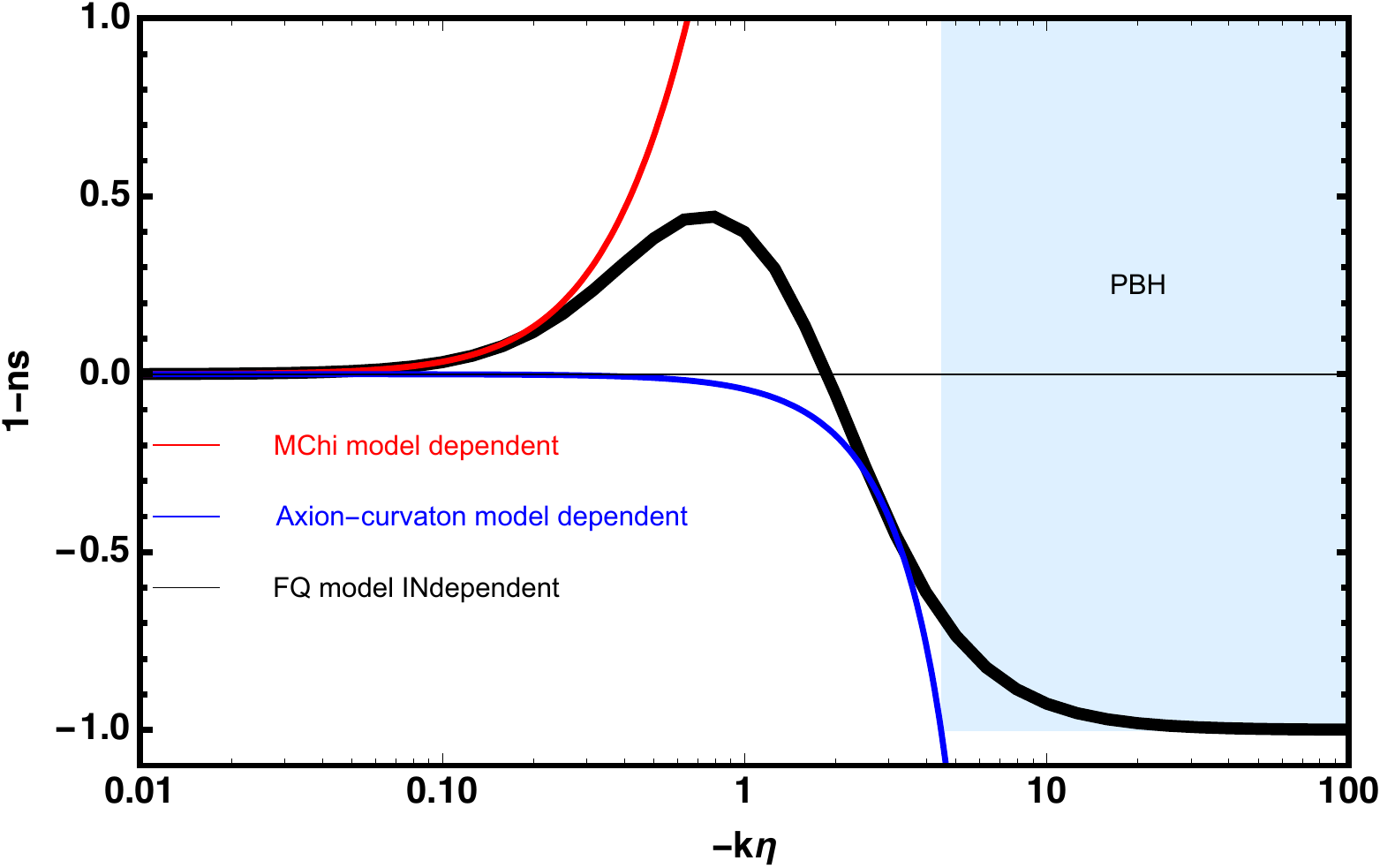} 
\caption{The black line shows the tilt $\frac{\alpha_F}{2}=1-n_s$ in the power spectrum of primordial fluctuations as a function of scale $k \eta$ computed from the, model independent, quantum Fisher ${\cal{F}}_Q$. There are several regimes. The nearly scale invariant regime $n_s -1 \sim 0.05$ corresponds to scales accessible by CMB+LSS observations. The regime with a blue tilt $n_s -1 = 1$ is the one we consider here; it corresponds to comoving scales smaller than 1 Mpc and is where PBH could be formed. We also show, for illustration, the typical red tilt of Starobinsky model as a function of scalaron scale  and the axion-curvaton blue tilt from~\ref{eq:axion} as a function of the axion scale $c$. The quantum tilt ( black line) defines a natural envelope of both regimes.} 
\label{fig:tilt}
\end{figure}

\section{Method and Results}

In Ref.~\cite{CR1}, that should be considered as a companion paper to this note, we have evaluated the quantum Fisher information for de Sitter vacua pure states. This quantum information measures the dependence of these states on the energy scale parameterizing the vacua (see also Ref.~\cite{GJ1,GJ2,GJ3} for applications of quantum Fisher to cosmology). The key finding in Ref.~\cite{CR1} was to show that this quantum Fisher information function has an anomalous transformation with respect to energy scale dilatations. This departure from scale invariance defines an energy scale dependent quantum tilt that is the one depicted in Fig.~\ref{fig:tilt}. Using the representation of the scale transformations in terms of the gauge invariant quantum oscillators, the purely quantum mechanical tilt derived for pure de Sitter states induces a tilt for the power spectrum of scalar fluctuations. Fixing the standard CMB scale $k\sim 0.005$ Mpc$^{-1}$ at the Fisher scale $k\eta=0.1$ where by definition $k\eta$ is the argument on which the Fisher quantum tilt depends sets the value of the spectral index to be $(1-n_s) = 0.00328$ \footnote{As discussed in \cite{GJ6} the selection of this point is not an arbitrary fit but the result of a self consistency relation. This crucial aspect of our construction will not be relevant for the rest of the discussion, so we address the interested reader to check the argument in \cite{GJ6}}  The important bonus of this quantum construction is that it provides non-perturbative information of the tilt on the whole range of scales. The quantum Fisher tilt derived in \cite{CR1} and depicted in Fig.~\ref{fig:tilt} contains four very different regimes, namely

\begin{enumerate}
\item An IR regime for values of the Fisher scale $(k\eta < 0.1)$. This regime ends at some IR scale $(k\eta)_{\rm end}$ where the quantum tilt starts increasing. This IR scale, that qualitatively set the end of the inflationary period, depends, in the numerical analysis we are developing, on the effective dimension of the Fock space of entangled pairs contributing to the quantum Fisher information ( see the sum in equation (3.2) of Ref.~\cite{GJ6}). A priori we don't have any physical way to set this number that we can send to infinity. The extension of the IR part in this limit grows until reaching $k\eta=0$ i.e. the inflation will not end. This is natural since we are not including any classical slow roll parameter in our construction. However, we can assume some sort of entropic limit to the number of entangled pairs contributing to the quantum Fisher. If we set this upper bound by the typical entropy $N_{GH}$ of the de Sitter space we are using, then the IR limit will correspond roughly to a Fisher scale $k\eta \sim \frac{1}{ \sqrt{N_{GH}}}$ that for an inflationary value of $H\sim 10^{15}$ Gev will corresponds to IR length scales of the order of $10^5$ Mpc. In essence the lesson of this IR part of the quantum tilt is to suggest a relation between the end of inflation and the maximal amount of pairs that can be created. It is important to stress that the UV behavior of the quantum tilt is, numerically, very insensible to the number of pairs contributing to the quantum Fisher.
\item A second regime of typical CMB+LSS  $k$ scales. In this regime, quantum effects are dominant and give rise to a quantum red tilt. We will denote this regime the CMB regime. In this regime the scale dependence of the power spectrum is determined by the red tilt of order $n_s=0.96$.
\item A transition region around horizon exit scales where the cosmological tilt transits from red into blue. In this transient region the spectral index $n_s$ reaches a minimal red value and starts increasing until becoming blue and bigger than one. The scale at which this transition takes place can be derived from the pivot Fisher scale $(k\eta) \sim 0.1$ corresponding to the CMB scale. In these conditions the transit into the blue phase takes place around scales $O(1 {\rm Mpc}^{-1})$.
\item Finally we have an UV regime of large values of $k$ characterized by a blue tilt with $n_s$ increasing from $1$ to $2$ at scales $k\eta$ larger than $10$.
\end{enumerate}

This quantum Fisher phase diagram (see Fig.~\ref{fig:tilt}) clearly shows the existence of the two needed regimens to produce PBH's. Indeed, the power spectrum will go from the small value observed in CMB scales to a larger value in the UV regime for large values of $k$. The simplest way to parameterize this value is setting the power spectrum in the transient region to be the CMB value $O(10^{-9})$ and to run this value in the blue tilt phase, namely
\begin{equation}\label{two}
{\cal{P}}_{\rm UV}(k) \sim 10^{-9} (\frac{k}{k_c})^{n_s(k)-1}
\end{equation}
with $k_c$  a pivot scale defining the transient region where the nature of the tilt changes from red to blue and with $n_s(k)$ defined in the UV region i.e. $n_s>1$ (See Fig.~\ref{fig:tilt} for the dependence of the quantum tilt on scales). In curvaton-like models \cite{Kawasaki:2012wr} the blue tilt is defined in terms of the curvaton scale $c$ ( $V_{cur} \sim cH^2\phi^2$ ) as
\begin{equation}
n_s-1 = 3 - 3\sqrt{1-\frac{4c}{9}}
\label{eq:axion}
\end{equation}
and $k_c\sim \frac{1}{\rm Mpc}$.
To make contact with the quantum tilt depicted in Fig.~\ref{fig:tilt} we take a CMB pivot scale $(k\eta)_{CMB} \sim 0.1$ with the transition region at $(k\eta)\sim 1$. The quantum tilt for $(k\eta) >1$ goes to $n_s=2$, so in order to create a power spectrum with value of order $10^{-3}$, which is  what is needed to generate PBH with the critical density (see Ref.~\cite{Kawasaki:2012wr}), with this $n_s=2$ tilt we need a Fisher scale
\begin{equation}
(k\eta) > 10^{6}
\end{equation}
i.e scales smaller than $10^{-6}$ Mpc.

This naturally leads, if we require $\Omega_{CDM} \sim \Omega_{PBH}$, to PBHs of mass around $10^{20}gr$ and smaller. It is worth noting that this mass range for PBHs is perfectly allowed experimentally (see Fig.~1 in Ref.\cite{CarrReview}).

In order to highlight the relation with inflationary model predictions we show, for illustrative purposes in the lower panel of Fig.~\ref{fig:tilt} the typical inflaton red tilt ~\cite{Mukhanov}  in red as well as the axion curvaton blue tilt \ref{eq:axion} in blue (where we use the scale dependent parameterization $c = 0.25 (k \eta)^2$). This illustrates how the quantum tilt defines a formal envelope of both types of behavior: namely the red one at small momentum scales we expect in Starobinsky model and the blue tilt at large momentum scales typical of curvaton like models.

Regardless of how much realistic the previous estimates can be, what we find quite remarkable of the former discussion is that  purely quantum mechanical properties of de Sitter vacua {\it provide naturally and in a completely model independent way the blue tilt in the UV regime of large values of $k$ required to trigger the PBH formation needed to identify them with dark matter.}

It is worth commenting on the slight ``bump" at horizon crossing scales $k \eta \sim 1$ that can be seen in Fig.~\ref{fig:tilt}. This will correspond to scales of tens of Mpc and will translate into a lack of power with respect to the CMB power spectrum because the spectrum is slightly redder. It is interesting that this could be related to the current ``$\sigma_8$ tension"~\cite{KIDS} seen in weak-lensing surveys and that translates into a lack of power at $\sim 10$ Mpc scales with respect to the one inferred using the CMB value in the LCDM model. We will investigate this in detail, as well as the super-horizon scales, in the future.

Before ending this note let us make a rather speculative comment on the potential microscopic meaning of the change in behavior of the tilt from red to blue at small scales. Inspired by the blue tilt of the axion curvaton model we could imagine that the quantum estimator naturally associated with our quantum Fisher information changes from an inflaton like field in the red tilted regime to a Peccei-Quinn axion like estimator in the blue regime (see Ref.~\cite{Cesar} for the relation of axions and quantum Fisher) where the energy scale dependence is related to the dependence on a CP breaking $\theta$ parameter. 

\section{Summary}

In summary, the message of this note is to point out to a very precisely defined quantum window for the origin of dark matter in the form of a red-blue transition of the quantum tilt that can trigger PBH formation. In essence, from this point of view: {\it Dark matter can be thought as just a model independent consequence of the quantum conformal properties of pure de Sitter as encoded in the scale dependence of the quantum tilt.} Independently of the implications on PBH formation, the prediction on the quantum tilt at small length scales is model independent and thus should be experimentally falsifiable with CMB spectral distortion observatories that probe these small scales (see e.g. Ref.~\cite{Chluba,Kogut}).

\acknowledgments
The work of CG was supported by grants SEV-2016-0597, FPA2015-65480-P and PGC2018-095976-B-C21. The work of RJ is supported by MINECO grant PGC2018-098866-B-I00 FEDER, UE.

\end{document}